\documentclass[aps,amsfonts,prl,showpacs,nobibnotes,nofootinbib,%
tightenlines]{revtex4}
\usepackage{amsmath,amssymb}
\input BoxedEPS
\SetTexturesEPSFSpecial  
\HideDisplacementBoxes

\newcommand{\half}{\frac{1}{2}}

\newcommand{\beq}{\begin{equation}}
\newcommand{\eeq}{\end{equation}}
\newcommand{\ba}{\begin{array}}
\newcommand{\ea}{\end{array}}

\begin{document}

\title{The Width of the $\Theta^{+}$ Exotic Baryon in the Chiral 
Soliton Model}
\author{R.~L.~Jaffe} 
 
\affiliation{Center for Theoretical Physics, Laboratory for Nuclear
Science and Department of Physics, Massachusetts Institute of Technology,
Cambridge, Massachusetts 02139\\
MIT-CTP-3470} 

\begin{abstract}\noindent 
    
In 1997 Diakonov, Petrov, and Polyakov, calculated the width of the
exotic baryon that they called $\Theta^{+}$.  The
prediction, $\Gamma(\Theta^{+})\lesssim$ 15 MeV, has received
considerable attention, especially in light of the narrowness of the
experimentally reported $\Theta^{+}$ resonance.  However, there is an
arithmetic error in their work: when corrected, the width
estimate quoted in that paper should have been $\lesssim$~30~MeV.
\end{abstract} 
\vspace*{-\bigskipamount}

\preprint{MIT-CTP-3470}
\maketitle

The existence of a relatively light exotic antidecuplet of baryons was
first pointed in the context of the Skyrme model by Manohar in
1984\cite{Manohar,Chemtob,BandD}\footnote{Diakonov, Petrov and
Polyakov\cite{dpp} cite Ref.~\cite{DP} in this regard.  However there
is no mention of the antidecuplet in either of the quoted
papers\cite{VK}.} The mass of the lightest and potentially most
prominent member of the antidecuplet, now known as the $\Theta^{+}$,
was computed by Praszalowicz in 1987\cite{mp}.  He predicted that this
$I=0$, $Y=2$ $K^{+}n$ resonance would have a mass of approximately
1530 MeV.

In a 1997 paper remarkable for its foresight, Diakonov, Petrov and
Polyakov studied the antidecuplet in the chiral soliton
model\cite{dpp}.  They obtained a mass estimate close to
Praszalowicz's and observed that experiments had not yet probed this
region thoroughly.  Their paper stimulated an experimental search for
the $\Theta^{+}$.  They also presented qualitative arguments that the
$\Theta^{+}$ could be quite narrow.  It is clear from their paper that
they believed this independent of any specific model calculation. 
However they present, and quote in the abstract and conclusions, a
specific calculation of the width of the $\Theta^{+}$ and other
antidecuplet baryons.  However the 15 MeV width they quote for the
$\Theta^{+}$ cannot be obtained from their equations.  Instead,
evaluation of their equations yield $\Gamma(\Theta^{+}\to NK)\approx
30$ MeV. The error is arithmetic.  Their model is simple, and clearly
and consistently presented.  One parameter ($G_{0}+\frac{1}{2}G_{1}$)
is fit to the width of the $\Delta(1232)$, the other ($G_{1}/G_{0}$)
is taken from chiral quark soliton models.  The results do not follow
from the numbers.

The purpose of this short note is to correct the arithmetic in
Ref.~\cite{dpp}.  Normally this would not require publication. 
However, three considerations motivate broad distribution of this
Comment: First, of course, the $\Theta^{+}$ has been discovered and
appears to be very narrow\cite{Nakano:2003qx}; second, the
``prediction'' of $\Gamma(\Theta^{+})\approx 15$ MeV in
Ref.~\cite{dpp} is frequently and prominently cited as an explanation
of the observed width; and finally, the authors of Ref.~\cite{dpp}
have declined to provide an erratum to their paper that would correct the
misperception propagated by this error.\footnote{The error in
Ref.~\cite{dpp} appears to have been recognized immediately after
publication of Ref.~\cite{dpp} by H.~Weigel in Ref.~\cite{hw}.  He
discusses the issue in a footnote on page 17 of the e-print.}

There is no physics at issue, so it is not necessary to review the
authors' model or the method by which they compute widths.  The issue
is arithmetic.  Ref.~\cite{dpp} presents separate expressions for the
partial widths for $\Delta\to N\pi$ (eq.~(42)),
$\Sigma^{*}\to\Lambda\pi$ (eq.~(43)), $\Sigma^{*}\to\Sigma\pi$
(eq.~(44)), $\Xi^{*}\to\Xi\pi$ (eq.~(45)), and $\Theta^{+}\to NK$
(eq.~(56)), and for other antidecuplet decays (eqs.~(57--67)).  All of
these equations are consistent with and summarized by their eq.~(49),
``the \ldots general formula for partial widths of members of the
decuplet and of the antidecuplet''\cite{dpp},
\begin{equation}
    \Gamma(B_{1}\to B_{2}M) = \frac{3G_{r}^{2}}{2\pi(M_{1}+M_{2})^{2}}
    |\vec p|^{3}\frac{M_{2}}{M_{1}}\left( 
    C_{1}+\frac{1}{\sqrt{5}}C_{2}c_{\overline{10}}\right)
    \label{49}
\end{equation}
Here $M_{1}$ and $M_{2}$ are the masses of baryons $B_{1}$ and $B_{2}$ 
respectively.  $|\vec p|$ is the center of mass momentum in the decay,
$$
|\vec p| = \frac{1}{2M_{1}}\sqrt{M_{1}^{4}+M_{2}^{4}+m^{4}
-2m^{2}M_{1}^{2} -2M_{1}^{2}M_{2}^{2} -2M_{2}^{2}m^{2}}.
$$
$m$ is the meson mass.  $G_{r}$ is a sum of Yukawa coupling constants
which differs for an initial decuplet, $r={10}$ or antidecuplet,
$r=\overline{10}$,
\begin{align}
    G_{10}& = G_{0}+\frac{1}{2}G_{1}\nonumber\\
    G_{\overline{10}} &= G_{0}-G_{1}-\frac{1}{2}G_{2}
\end{align}
as listed in Table 2 of Ref.~\cite{dpp}.  The authors take
$G_{2}\approx 0$ and $c_{\overline{10}}\approx 0$ in their numerical
evaluation of widths.  The constant $C_{1}$ is a Clebsch-Gordan
coefficient.  It equals 1/5 for both $\Delta\to N\pi$ and
$\Theta^{+}\to NK$, which are the decays relevant here.  To compute
decuplet and antidecuplet widths, it is necessary to know both $G_{0}$
and $G_{1}$.  The authors fit $G_{0}+\frac{1}{2}G_{1}$ to the width of
the $\Delta(1232)$ and use the chiral quark soliton model to estimate
$G_{1}/G_{0}\approx 0.4$.

The crux of the issue is the application of eq.~(\ref{49}) to the 
$\Delta$ and $\Theta^{+}$ decays.  The authors provide the formulas 
for each case.  Eq.~(42) applies to the $\Delta$,
\begin{equation}
    \Gamma(\Delta\to N\pi) =\frac{3(G_{0}+\half G_{1})^{2}}{2\pi
    (M_{\Delta}+M_{N})^{2}}|\vec 
    p|^{3}\frac{M_{N}}{M_{\Delta}}\frac{1}{5},
    \label{42} 
\end{equation}
where, following Ref.~\cite{dpp} I have replaced $G_{0}$ by 
$G_{0}+\half G_{1}$.  Eq.~(56) applies to the $\Theta$,
\begin{equation}
    \Gamma(\Theta\to NK) =\frac{3(G_{0}-G_{1})^{2}}{2\pi
    (M_{\Theta}+M_{N})^{2}}|\vec 
    p|^{3}\frac{M_{N}}{M_{\Theta}}\frac{1}{5},
   \label{56}
\end{equation}
where, following Ref.~\cite{dpp} I have dropped a term proportional to 
$G_{2}$ and another proportional to $c_{\overline {10}}$.  Substituting known 
masses and $\Gamma(\Delta\to N\pi)=110$~MeV into eq.~(\ref{42}) one 
finds,
\begin{equation}
    G_{0}+\frac{1}{2}G_{1}\approx 25
    \label{corr}
\end{equation}
in disagreement with the value $G_{0}+\frac{1}{2}G_{1}\approx 19$
quoted in eq.~(54) of Ref.~\cite{dpp}.  This appears to be an
arithmetic error.\footnote{In Ref.~\cite{hw} Weigel pointed out that
the discrepancy would be explained if the authors of Ref.~\cite{dpp}
had mistakenly used $M_{\Delta}/M_{N}$ in place of $M_{N}/M_{\Delta}$
in eq.~(\ref{42}).  This error was subsequently acknowledged in
correspondence between Weigel and one of the authors of
Ref.~\cite{dpp}.\cite{hwthanks}} Substitution of the correct value, eq.~(\ref{corr}),
into eq.~(\ref{56}) along with $G_{1}/G_{0}\approx 0.4$\cite{dpp} gives
\begin{equation}
\Gamma(\Theta^{+}\to NK)\approx 30 \ \mbox{MeV}
\label{theta}
\end{equation}
in disagreement with the result $\Gamma(\Theta^{+}\to NK)= 15$ MeV 
claimed in Ref.~\cite{dpp}.  

It is also worth noting that the arithmetic error afflicts the value of 
the $\pi N$ coupling constant, $g_{\pi N N}$, in the model of
Ref.~\cite{dpp}.  Instead of $g_{\pi N N}\approx 13.3$ as claimed in 
eq.~(54) of Ref.~\cite{dpp}, the corrected result is
$$
g_{\pi NN}=\frac{7}{10}(G_{0}+\frac{1}{2}G_{1})\approx 17.5
$$
compared with the experimental value of $\approx$ 13.6.  If
$G_{0}+\frac{1}{2}G_{1}$ is adjusted to obtain the correct value of
$g_{\pi NN}$, then the model prediction of the width of the $\Delta$ is
too small by almost a factor of two.  The arithmetic error generates
minor corrections to the widths of the decuplet baryons listed in
eqs.~(42)--(45).  It also modifies the predicted widths of the other
antidecuplet states, increasing all of them by roughly a factor of
two.

To summarize, the purpose of this short note has been to point out and
correct an arithmetic error in Ref.~\cite{dpp}.  Since arithmetic is
performed according to universal rules, and is considerably simpler
than theoretical physics, the reader should feel free to check the
issue for him or herself directly.  

\subsection{Acknowledgments}
I thank H.~Weigel for correspondence on this issue.  I also 
acknowledge correspondence with D.~Diakonov and M.~Polyakov.
This work is supported in part by the U.S.~Department of Energy
(D.O.E.) under cooperative research agreement~\#DF-FC02-94ER40818.



\end{document}